\documentstyle[twocolumn,prl,aps,epsfig]{revtex}
\begin{document}                                                              
\newcommand{\be}{\begin{eqnarray}}
\newcommand{\ee}{\end{eqnarray}}
\newcommand{\dd}{\mathrm{d\,}}
\title{The onset of classical QCD dynamics in relativistic heavy ion collisions}
\author{Dmitri Kharzeev$^a$, Eugene Levin$^b$ and Marzia Nardi$^c$}

\bigskip
 
\address{
a) Department of Physics,\\ Brookhaven National Laboratory,\\
Upton, New York 11973-5000, USA\\
b) HEP Department, School of Physics,\\
Raymond and Beverly Sackler Faculty of Exact Science,\\
Tel Aviv University, Tel Aviv 69978, Israel\\
c) Dipartimento di Fisica Teorica dell'Universit{\`a} di Torino and INFN, Sezione di Torino\\ 
via P.Giuria 1, 10125 Torino,
     Italy}
\date{November 23, 2001}
\maketitle
\begin{abstract} 

The experimental results on hadron production obtained recently at RHIC offer a new prospective on 
the energy dependence of the nuclear collision dynamics. In particular, it is possible that 
parton saturation -- the phenomenon likely providing initial conditions for the multi--particle production at RHIC 
energies -- may have started to set in central heavy ion collisions already around the highest SPS energy. 
We examine this scenario, and make predictions based on high density QCD for the 
forthcoming $\sqrt{s}=22$ GeV run at RHIC.  
\end{abstract}
\begin{narrowtext}


High energy nuclear 
collisions allow to test QCD   
at the high parton density, strong color field frontier, where 
the highly non--linear behavior is expected. 
Already after one year of RHIC operation, a wealth of new experimental information on 
multi-particle production has become available \cite{Phobos,Phenix,Star,Brahms}. It appears that    
the data on hadron multiplicity and its energy, centrality and rapidity dependence 
so far are consistent with the approach \cite{KN,KL} based on the ideas 
of parton saturation \cite{GLR,hdQCD} and semi--classical QCD (``the color glass condensate'') \cite{MV,MV1}. 
The centrality dependence of transverse mass spectra appears to be consistent with this 
scenario as well \cite{SB}.  

Strictly speaking, the use of classical weak coupling methods in QCD can 
be justified only when the ``saturation scale'' $Q_s^2$ \cite{GLR,MV}, proportional to the 
density of the partons, 
becomes very large, $Q_s^2 \gg \Lambda^2_{\rm QCD}$ and 
$\alpha_s(Q_s^2) \ll 1$. 
At RHIC energies, the value 
of saturation scale in $Au-Au$ collisions varies in the range of $Q_s^2 = 1 \div 2\ 
{\rm GeV}^2$ depending on centrality. At these values of $Q_s^2$, we are still 
at the borderline of the weak coupling regime. However, the apparent success 
of the saturation approach seems to imply that the transition to semi--classical QCD dynamics 
takes place already at RHIC energies.
      
This may shed new light on the mechanism of hadron production at lower energies, 
perhaps including the energy of CERN SPS. Indeed, extrapolating down in energy using 
the formulae of \cite{KL} yields for saturation scale in 
central $Pb-Pb$ collisions at SPS energy of $\sqrt{s}=17$ GeV the value of 
$Q_s^2 \approx 1.2\ {\rm GeV}^2$ \footnote{We use $Q^2_s \propto 
s^{\lambda/2}$ with $\lambda \approx 0.25 \div 0.3 $ as it follows from the  
scaling behavior of the HERA data \cite{GW}; see below.}  . The same average value at a 
RHIC energy of $\sqrt{s}=130$ GeV is reached in peripheral $Au-Au$ 
collisions at impact parameter $b \approx 9$ fm and an average number of 
participants of $N_{part} \approx 90$. At $N_{part} < 100$, and impact parameters $b > 9$ fm, 
reconstruction of the geometry of the collision and the extraction of the number of participants 
face sizable uncertainties, and no firm conclusion on the applicability of the saturation 
approach can be drawn from the data. Given this uncertainty, one may consider two different 
scenarios: 

1) the onset of saturation occurs somewhere in the RHIC energy range, 
below $\sqrt{s}=130$ GeV but above $\sqrt{s}=17$ GeV; 
the mechanisms of multi--particle production at RHIC and SPS 
energies are thus totally different; 

2) saturation sets in central heavy ion collisions 
already around the highest SPS energy. The second scenario would, in particular, have important 
implications for interpretation of the SPS results.

It should be possible to distinguish between these two scenarios by extrapolating the 
results of Refs.\cite{KN,KL} down in energy and comparing them to the data. 
In fact, very soon RHIC will collect data at the 
energy of $\sqrt{s}=22$ GeV, not far from the highest SPS energy of $\sqrt{s}=17$ GeV. In this Letter, we make 
predictions for hadron production at this energy based on the saturation scenario. It should be stressed   
that {\it a priori} there is no solid reason to expect this approach to work at low energies; 
we provide these predictions to make it possible to decide between scenarios 1) and 2) 
listed above basing on the data when they become available.

In Ref.\cite{KL} we derived a simple analytical scaling formula, describing 
the energy, centrality, rapidity, and atomic number dependences of hadron multiplicities 
in high energy nuclear collisions:
\be
{dN \over d y} = c\ N_{part}\ \left({s \over s_0}\right)^{\lambda \over 2}\ e^{- \lambda |y|}\ 
\left[\ln\left({Q_s^2 \over \Lambda_{QCD}^2}\right) - \lambda |y|\right]\ \times \nonumber
\ee
\be
\times \left[ 1 +  \lambda |y| \left( 1 - {Q_s \over \sqrt{s}}\ e^{(1 + \lambda/2) |y|} \right)^4 \right],  
\label{finres}
\ee
with $Q_s^2(s) = Q_s^2(s_0)\ (s /s_0)^{\lambda / 2}$.
Once the energy--independent constant $c \sim 1$ and $Q_s^2(s_0)$ are determined 
at some energy $s_0$, Eq. (\ref{finres}) contains no free parameters. (The value of $\lambda$, 
describing the growth 
of the gluon structure functions at small $x$ can be determined in deep--inelastic scattering; 
the HERA data are fitted with $\lambda \simeq 0.25 \div 0.3$ \cite{GW}).
At $y = 0$ the expression (\ref{finres}) coincides with the one 
derived in \cite{KN}, and extends it to describe the rapidity and energy dependences. 

Using the value of $Q_s^2 \simeq 2.05\ {\rm GeV}^2$ extracted \cite{KN} at $\sqrt{s} = 130$ GeV and $\lambda = 0.25$ 
\cite{GW} used in \cite{KL}, equation (\ref{finres}) leads to the following approximate formula  
for the energy dependence of charged multiplicity in central $Au-Au$ collisions:
\be
\left<{2 \over N_{part}}\ {d N_{ch} \over d \eta}\right>_{\eta < 1} \approx 0.87\ 
\left({\sqrt{s}\ ({\rm GeV}) \over 130}\right)^{0.25}\ \times \nonumber
\ee
\be
\times \left[3.93 + 0.25\ \ln\left({\sqrt{s}\ ({\rm GeV}) \over 130}\right)
\right]. \label{endep}
\ee
At $\sqrt{s} = 130\ {\rm GeV}$, we estimate from Eq.(\ref{endep}) 
$2/N_{part}\ dN_{ch}/d\eta \mid_{\eta<1} = 3.42 \pm 0.15$, 
to be compared to the average experimental value of $2/N_{part}\ dN_{ch}/d\eta \mid_{\eta<1} = 3.37 \pm 0.12$ 
\cite{Phobos,Phenix,Star,Brahms}. 
At $\sqrt{s} = 200\ {\rm GeV}$, one gets $2/N_{part}\ dN_{ch}/d\eta \mid_{\eta<1} = 3.91 \pm 0.15$, 
to be compared to the PHOBOS value \cite{Phobos} of $2/N_{part}\ dN_{ch}/d\eta \mid_{\eta<1} = 3.78 \pm 0.25$. 
Finally, at $\sqrt{s} = 56\ {\rm GeV}$, we find $2/N_{part}\ dN_{ch}/d\eta \mid_{\eta<1} = 2.62 \pm 0.15$, 
to be compared to \cite{Phobos} $2/N_{part}\ dN_{ch}/d\eta \mid_{\eta<1} = 2.47 \pm 0.25$. 
Having convinced ourselves that our result (\ref{endep}) describes the experimentally observed 
energy dependence of hadron multiplicity in the entire interval of existing measurements at RHIC 
within error bars, we can extrapolate it to the small energy of $\sqrt{s} = 22\ {\rm GeV}$ 
and make a prediction:
\be
\left<{2 \over N_{part}}\ {d N_{ch} \over d \eta}\right>_{\eta < 1} = 1.95 \pm 0.1; \ \ \  \sqrt{s} = 22\ {\rm GeV}.
\ee

It is also interesting to note that formula (\ref{endep}), when extrapolated to very high energies, 
predicts for the LHC energy a value 
substantially smaller than found in other approaches:
\be
\left<{2 \over N_{part}}\ {d N_{ch} \over d \eta}\right>_{\eta < 1} = 10.8 \pm 0.5; \ \ \  \sqrt{s} = 5500\ {\rm GeV},
\ee
corresponding only to a factor of $2.8$ increase in multiplicity between the RHIC energy of  $\sqrt{s} = 200\ {\rm GeV}$ 
and the LHC energy of $\sqrt{s} = 5500\ {\rm GeV}$  
(numerical calculations show that when normalized to the number of participants, 
the multiplicity in central $Au-Au$ and $Pb-Pb$ systems is almost identical).

Let us now turn to the centrality dependence.   
Our method has been described in detail before \cite{KN,KL}. 
We first use Glauber approach to reconstruct geometry of the collision, and then apply 
semi--classical QCD to evaluate the multiplicity of produced gluons at a given 
centrality and pseudo--rapidity.  
The Glauber formalism (see \cite{KN,KLNS} for details) allows to evaluate 
the differential cross of inelastic nucleus--nucleus 
interaction at a given (pseudo)rapidity $\eta$:

\be
\frac {d \sigma} {d n} = \int d^2b \ {\cal P}(n;b)\ (1 - P_0(b));  \label{dsig}
\ee
$P_0(b)$ is the probability of no interaction among the nuclei at a given 
impact parameter $b$: 
\be
P_0(b) = (1 - \sigma_{NN} T_{AB}(b))^{AB}, 
\ee
where  
$\sigma_{NN}$ is the inelastic nucleon--nucleon cross section, and $T_{AB}(b)$ is the 
nuclear overlap function for the collision of nuclei with atomic numbers A and B; 
we have used the three--parameter Woods--Saxon nuclear density distributions \cite{tables}.
For $\sqrt{s} = 22\ {\rm GeV}$ we use $\sigma_{NN}=33 \pm 1 \ {\rm mb}$ 
basing on the interpolation of existing $pp$ data 
\cite{PDG}.   
The correlation function ${\cal P}(n;b)$ has a Gaussian form described in \cite{KLNS,KN}. 
The total cross section of inelastic hadronic {\it Au--Au} interactions 
computed in our approach at $\sqrt{s}=22$ A GeV is 
$\sigma_{tot} = 6.9 \pm 0.05$ barn.

The correspondence between a given centrality cut and the mean numbers of nucleon 
participants and nucleon--nucleon collisions 
can now be established by computing the average over the distribution (\ref{dsig}), as 
described in \cite{KN}. At  $\sqrt{s} = 22\ {\rm GeV}$ for $Au-Au$ collisions we 
find for the $0-6\%$ centrality cut $\left< N_{part} \right> = 332 \pm 2$;  
 $\left< N_{coll} \right> = 828 \pm 6$.
For $15-25\%$ centrality cut, we have $\left< N_{part} \right> = 179 \pm 2$ and 
 $\left< N_{coll} \right> = 367 \pm 6$, while for the $35-45\%$ cut, 
corresponding to rather peripheral collisions with 
the average impact parameter of $\langle b \rangle \approx 9.5$ fm, one gets
 $ \left< N_{part} \right> = 77 \pm 2$ and $\left< N_{coll} \right> = 120 \pm 5$.

We now have the information about the geometry of the collision needed to proceed with our 
calculation of centrality dependence of hadron multiplicities in the semi--classical QCD approach. 
However, in applying this method at small energies and/or for peripheral collisions, we face 
a fundamental dilemma.
In semi--classical approach, the multiplicity of the produced gluons is proportional to $1/\alpha_s(Q_s^2)$ 
(this is, of course, the origin of the factor $\ln(Q_s^2/\Lambda_{QCD}^2)$ in our formula (\ref{finres})). 
Once the saturation scale $Q_s^2$ becomes small, the result thus becomes sensitive to 
the behavior of the strong coupling in the infra--red region. Taking a conservative viewpoint, 
this simply signals that the method ceases to be applicable. 
If we accept this, we have to stop and conclude in favor of scenario 1) described above.  

However, this is not necessarily correct -- there is a solid body of evidence 
from jet physics that QCD coupling stays reasonably small, $\langle \alpha_s \rangle_{IR} = 0.4 \div 0.6$ 
in the infrared region \cite{YD}. The ``freezing'' at small virtualities solution for the QCD coupling 
$\langle \alpha_s \rangle_{IR} \approx 0.43$ has been found by Gribov \cite{VG} as a 
consequence of ``super--critical'' screening of color 
charge by light quark--antiquark pairs. Matching QCD onto the chiral theory through scale anomaly 
leads to the coupling frozen in the infrared region, with magnitude 
$\langle \alpha_s \rangle_{IR} \approx 0.56$ \cite{FK}. ``Freezing'' solutions for the running 
coupling are repeatedly discussed; see \cite{AS} for a recent review. 
It is possible that the presence of relatively large scale 
in hadro--production reflects the properties of QCD vacuum \cite{KL1,ES}. 
      
We will thus try to adopt an optimistic point of view and assume that the strong coupling indeed ``freezes'' 
below $Q_s^2 \approx 0.8 \ {\rm GeV}^2$ at the value of $\langle \alpha_{s}\rangle_{IR} \approx 0.5$. 
In fact, one may even dare to go further -- {\it assuming} the validity of semi--classical 
QCD approach at low energies, 
the centrality dependence of hadron multiplicity may be used to glean information about the behavior of 
 strong coupling in the infra--red region.  

\begin{figure}[b]
\epsfig{file=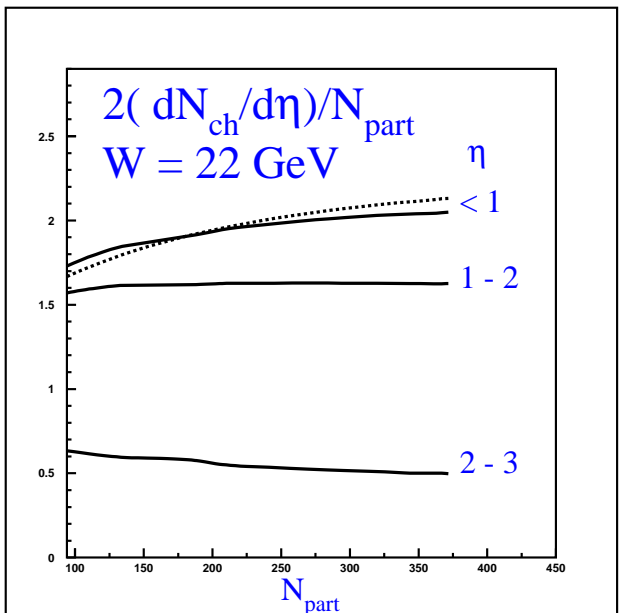, height=3.2in}
\vskip0.1cm
\caption{Centrality dependence of the charged multiplicity per participant pair at different 
pseudo--rapidity intervals at 
$\sqrt{s} = 22$ A GeV; see text for details.}
\label{centr}
\end{figure}

To evaluate the resulting centrality dependence around $\eta = 0$ we use two different ans{\"a}tze 
for the running coupling: a) ``smooth freezing'' $\alpha_s \sim 1/\ln((Q_s^2 + \Lambda^2)/\Lambda_{QCD}^2)$, with 
$\Lambda = 0.8\ {\rm GeV}^2$; b) ``sudden freezing'', when $\alpha_s$ is simply put equal to $\alpha_s(\Lambda^2)$ when 
$Q_s^2 < \Lambda^2$. The results are shown in Fig. \ref{centr} by solid (ansatz a)) and 
dashed (ansatz b)) lines. Note that even in the case of ``sudden freezing'', centrality dependence 
is smooth -- this is because the fraction of the transverse area where the local value of $Q_s^2$ becomes 
smaller than the cutoff $\Lambda^2$ is a smooth function of centrality.

Let us now discuss rapidity dependence. Unfortunately, we have found that at 
at low energies $\sqrt{s} \sim 20$ GeV the expression (\ref{finres}) 
provides a poor approximation to the numerical result based on the general formula \cite{GLR,GM} 
used in \cite{KL}:
\be
E {d \sigma \over d^3 p} = {4 \pi N_c \over N_c^2 - 1}\ {1 \over p_t^2}\  \times \nonumber
\ee
\be
\times \ \int d k_t^2 \ 
\alpha_s \ \varphi_A(x_1, k_t^2)\ \varphi_A(x_2, (p-k)_t^2), \label{gencross}    
\ee
where $x_{1,2} = (p_t/\sqrt{s}) \exp(\pm y)$ and 
$\varphi_A (x, k_t^2)$ is the unintegrated gluon distribution. 
This happens because at low energies the limited phase space suppresses the transverse momentum 
distribution of the produced gluons already below the saturation momentum $Q_s$. 
We thus have to evaluate the integral in (\ref{gencross}) numerically (see \cite{KL} for the list 
of formulae needed for this computation). 

\begin{figure}[b]
\epsfig{file=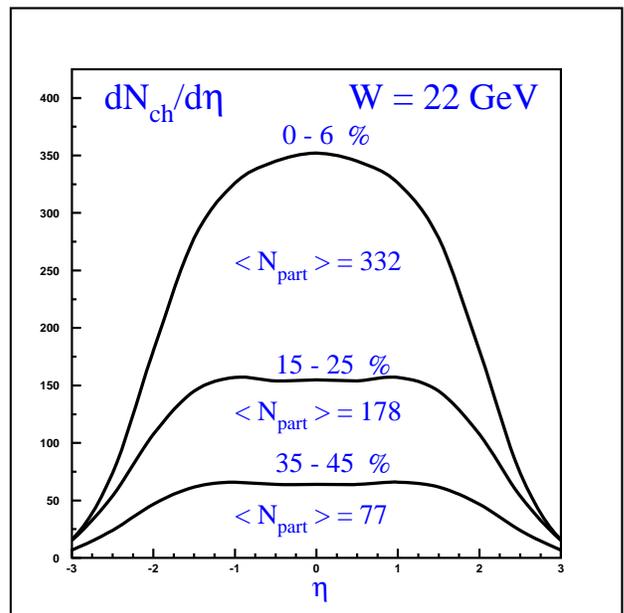, height=3.2in}
\vskip0.1cm
\caption{Pseudo--rapidity distribution of the charged multiplicity per participant pair in different 
centrality cuts at 
$\sqrt{s} = 22$ A GeV.}
\label{rap}
\end{figure}

To convert the computed rapidity distributions of gluons to the observed pseudo--rapidity distribution of hadrons, 
we follow the procedure of \cite{KL}, assuming that the ``local parton--hadron duality'' (see \cite{YD} and references 
therein) in the space 
of emission angles $\theta$, or, equivalently, in pseudo--rapidity $\eta = - \ln \tan (\theta/2)$. 
This corresponds to the physical assumption that once a gluon has been emitted along a certain direction, 
its final state interactions and fragmentation will not significantly change the direction of the resulting hadrons.   
The results of our calculations are presented in Figs. \ref{centr} and \ref{rap}. 
 
How reliable are our results, apart from the obvious leap of faith involved in the application 
of semi--classical method at small energy? The least reliable of our predictions is 
the distribution in pseudo--rapidity; indeed, we have assumed that multi--particle production is dominated 
by gluons, and this is not necessarily so at $\sqrt{s}=22$ GeV, where valence quarks may give essential 
contribution even at central rapidity. Also, since we are quite close to the fragmentation region at this 
energy, multi--parton correlation effects can also play a r{\^o}le. 
The absolute value of multiplicity around $\eta = 0$ is more stable, but still may  
be affected by the contribution from quarks and deviations from $\sim x^{- \lambda}$ behavior 
of the nuclear gluon distribution at larger $x$. 

There does exist however a prediction of our approach which is both robust and distinct: it is  
the rise with centrality of multiplicity per participant shown in Fig. \ref{centr} around $\eta \approx 0$. 
The shape of this dependence simply reflects the running of strong coupling, and is similar to the 
shape observed at much higher energy of 
$\sqrt{s} = 130$ GeV: indeed, the logarithm in Eq.(\ref{finres}) is a slowly varying function, and the dependence of 
saturation scale on energy is quite weak. This prediction is in marked contrast both to the 
strong increase of the slope of centrality dependence with energy predicted in a two--component model \cite{WG} 
(for a recent development, see however \cite{LW}) and 
to the final--state saturation model \cite{EKRT}, predicting a nearly constant, weakly decreasing,  
centrality dependence of multiplicity per participant.

\bigskip

The work of D.K. is supported by the U.S. Department of Energy under Contract No. DE-AC02-98CH10886. 
The research of E.L. was supported in part by the Israel Science 
Foundation, founded by the  Israeli Academy of science and 
Humanities, and BSF\, \# 9800276.

\end{narrowtext}
\end{document}